\documentstyle[prl,aps]{revtex}
\begin{document}
\draft
\title{
Instability of Anisotropic Fermi Surfaces in Two Dimensions}
\author{J. Gonz\'alez $^1$, F. Guinea $^2$ and M. A. H. Vozmediano $^3$ \\}
\address{
        $^1$Instituto de Estructura de la Materia. 
        Consejo Superior de Investigaciones Cient{\'\i}ficas. 
        Serrano 123, 28006 Madrid. Spain. \\
        $^2$Instituto de Ciencia de Materiales. 
        Consejo Superior de Investigaciones Cient{\'\i}ficas. 
        Cantoblanco. 28049 Madrid. Spain. \\
        $^3$Departamento de Matem\'aticas.
        Universidad Carlos III. 
        Butarque 15.
        Legan\'es. 28913 Madrid. Spain.}
\date{\today}
\maketitle
\begin{abstract}
The effect of strong anisotropy on the Fermi line of a system
of correlated electrons is studied in two space dimensions,
using renormalization group techniques. Inflection points change the
scaling exponents of the couplings, enhancing the instabilities of the
system. They increase the critical dimension for non Fermi
liquid behavior, from 1  to 3/2.
Assuming that, in the absence of nesting, the dominant
instability is towards a superconducting ground state, simple rules to
discern between d-wave and extended s-wave
symmetry of the order parameter are given. 

\end{abstract}

Since the appearance of high-T$_c$ materials 
the quest for non-Fermi liquid (NFL) behavior in two space
dimensions has attracted great interest. The emergence
of the quantum field theory  approach to the study of
electronic systems\cite{Shankar} seems to put severe constraints
on the possibility of NFL behavior, at least for continuum
models. Following Ref. \onlinecite{Shankar} 
no NFL behavior is to be
expected in D=2 provided that the coupling is weak, the
interaction is short-ranged, and the Fermi surface is isotropic.
On the other hand, Anderson\cite{And} and coworkers support strongly the
opinion that Fermi liquid behavior will not occur in D=2
under quite general assumptions, which make use of analogies to 1D
behavior. 
The basis of Anderson's
approach is the discrete 2D lattice. On the other hand, it has been
proved that the critical dimension for Luttinger liquid behavior in
isotropic systems is D = 1\cite{Dicastro}. 

This paper addresses
the role of anisotropy of the Fermi surface
in inducing instabilities of the Fermi liquid
fixed point in 2D.
The cases of an extreme
anisotropy due to either nesting or a Van Hove
singularity are known to produce drastic changes
in the system\cite{schulz,dzya,lee,mark,tsuei,vanhove}.
We study here the effect that a strong anisotropy of the Fermi line can
have in a continuum model in the absence of nesting and away
from a Van Hove singularity. 

In between the case of a circular Fermi surface 
and the existence of a Van Hove
singularity lies the possibility that the Fermi surface has 
inflection points which separate portions where the curvature has
opposite signs, as shown in Fig. \ref{ffermi}. This situation takes
place for a finite range of fillings, and requires no precise fine
tuning of the number of electrons in the system. In particular, for the
$t - t'$ Hubbard model in a square lattice, 
inflection points at the Fermi surface  are present for fillings below
the Van Hove point ( $- 8 t' + 16 t'^3 / t^2 < \varepsilon_F <
- 4 t'$, where $t < 0 , t' > 0$).

Scattering processes which involve two opposite inflection points 
are analogous to the enhanced scattering
present in nested Fermi surfaces\cite{Shankar}. It can be shown that
nesting, in any dimension, is a marginal perturbation. The correction
that nesting induces to the bare interaction (in dimensionless units)
is of order $\Lambda^0 \log \Lambda$, where $\Lambda$ is the 
high-energy cutoff\cite{Shankar}. A Van Hove singularity induces similar
effects in 2D\cite{vanhove}.

In order to study the effects induced by inflection points at the Fermi
surface, we first analyze the electron-hole polarizability
(the bubble in Fig. \ref{diagram}(a)) for a wavevector 
${\bf Q}$ which connects two such points
at opposite ends of the center of the Brillouin Zone (BZ). The imaginary part
of the diagram, ${\rm Im} \Pi ( {\bf Q} , \omega )$, measures the density of
electron-hole pairs with energy $\omega$.  In the case of perfect
nesting, each strip of holes of energy $\omega / 2$ below the Fermi
level coincides with a strip of electrons at energy $\omega / 2$ above
the Fermi level. Hence, ${\rm Im} \Pi ( {\bf Q} , \omega ) \sim | \omega |^0$,
and the corresponding coupling is marginal. Near an inflection point, the
states of equal energy no longer fill a straight strip in momentum
space.
Using a
reference frame such that the x-axis is parallel to the Fermi surface,
we can write:
\begin{eqnarray}
\varepsilon({\bf q}) & \approx   &v_F q_y + \alpha q_x^3 + \beta q_x^4 
\nonumber \\
\varepsilon({\bf q}+{\bf Q}) & \approx &-v_F q_y - \alpha q_x^3 + \beta q_x^4 
\label{disp1} 
\end{eqnarray}
The number of electron-hole pairs of energy $\omega$ available are given
by the region in momentum space which satisfy:
$\omega - d \omega \le \varepsilon_e ( {\bf q} ) - \varepsilon_h ( {\bf q +
Q} ) \le \omega$.

Setting $q_y \approx \omega / (2 v_F )$, these
restrictions imply that the region available is bounded, in the
transverse direction, by $q_x \propto q_y^{1/4} \propto \omega^{1/4}$
(note that the cubic term does not break the correspondence between the
equal energy slices at ${\bf q}$ and ${\bf q + Q}$).
This implies that ${\rm Im} \Pi ( {\bf Q}, \omega ) \propto
| \omega |^{1/4}$. The previous argument can easily be extended to an
arbitrary number of dimensions. An inflection point is characterized by a
dispersion relation which is linear in the direction normal to the Fermi
surface, cubic in one of the transverse directions, and quadratic in
all the others. In $D$ dimensions, this leads to:
\begin{equation}
{\rm Im} \Pi ( {\bf Q} , \omega ) \propto | \omega |^{\frac{D - 2}{2} +
\frac{1}{4}}
\label{scaling}
\end{equation}
Hence, the critical dimension is $D_c = 3/2$.
The same argument, when applied to the standard $2 k_F$ scattering which
connects two opposite points of an isotropic Fermi surface, gives ${\rm
Im} \Pi ( 2 k_F , \omega ) \propto | \omega |^{( D - 1 ) / 2}$, and the
corresponding critical dimension is $D_c = 1$, in agreement
with Ref. \onlinecite{Dicastro}.

The preceding analysis shows that inflection points enhance the
instabilities of the model.  Among the possible broken symmetry ground
states, we will assume that superconductivity prevails. The model has no
tendency towards ferromagnetism, as the density of states needs not be
specially large. A spin density wave, or a charge density wave, of
momentum ${\bf Q}$ is possible. The associated gap, however, will only
span a small fraction of the Fermi surface, centered around the
inflection points. Moreover, only two of the eight 
possible points of a square lattice will
take part in the symmetry breaking. As it is shown below, the system is
unstable against anisotropic superconductivity. This instability opens a
gap over large regions of the Fermi surface (except at the inflection
points). Thus, this instability is a more robust alternative than a spin or
a charge density wave.

The possible processes which renormalize the particle-particle
scattering are shown in Fig. \ref{diagram}. In a 3D isotropic system
with repulsive interactions, they lead to the Kohn-Luttinger
instability\cite{KL}, which gives rise to superconductivity at high angular
momenta.  Note that this instability is absent in 2D isotropic
systems\cite{kagan}.
For short-range interactions, the first and second diagrams in
Fig. \ref{diagram} cancel.  This fact implies that pairing, if present,
cannot arise from the propagator of a quasiparticle, in a
similar fashion to phonon mediated pairing. In that case, 
superconductivity can be expressed in terms of a diagram like the first
one
in Fig. \ref{diagram}. If the electrons to be paired and the
electrons which give rise to the quasiparticle are the same,  diagrams
like the first one in Fig. \ref{diagram} are cancelled, to all orders, by
diagrams like the following one in the same figure.

In general, we need to consider the scattering of a pair of electrons of
momenta ${\bf k} , {\bf - k}$ into ${\bf p} , {\bf - p}$. We label these
processes by the angle $\phi$ between ${\bf k}$ and the x-axis, and 
${\bf p}$ and the x-axis, $\phi '$. 
In the following, we will label each point at the Fermi surface by the
angular variable, $\phi$.
The value of the corresponding diagram in Fig. \ref{diagram}(c)
is denoted $\Pi ( \phi , \phi ' )$.

We first analyze contributions of the type $\Pi ( \phi , \phi )$. 
We assume a
local electron-electron interaction, $U$. Then, using
Eq. (\ref{disp1}) we have
\begin{equation}
 \Pi ( \phi , \phi )   =    -i
\frac{U^2}{(2\pi)^3} \int d\omega \int d^2 q 
e^{-\varepsilon ({\bf q}) / \Lambda}
e^{-\varepsilon ({\bf q} + {\bf Q} + {\bf \delta Q}) / \Lambda}
          G(\omega, {\bf q})
G(\omega, {\bf q} + {\bf Q} + {\bf \delta Q} )  
\label{vertexeq}
\end{equation}
where ${\bf Q} + {\bf \delta Q}$ is the vector which connects the point
at the Fermi surface with momentum ${\bf k}$ with that at $- {\bf k}$.
The parameter $\Lambda$ is a smooth cutoff
in energies which is used to implement the renormalization group 
(RG) scheme.

Using the new variables:
$ \varepsilon  =  \varepsilon ({\bf q} - {\bf \delta Q}/2 ) $ and
$ \overline{\varepsilon }  = \varepsilon ({\bf q} + {\bf Q}
  + {\bf  \delta Q}/2 ) $,
the real part of $ \Pi (\phi, \phi )$ becomes, for the convex
part of the Fermi line,
\begin{eqnarray}
{\rm Re} \Pi (\phi, \phi ) & = &   -\frac{U^2}{(2\pi^2)}
 \int_{0}^{\infty} d\varepsilon
  \int_{0}^{\infty} d \overline{\varepsilon} \left| \frac{
 \partial (q_x, q_y)}{ \partial (\varepsilon, 
    \overline{\varepsilon} ) } \right|
  \frac{e^{-\varepsilon / \Lambda} 
      e^{- \overline{\varepsilon} / \Lambda} }
   { \varepsilon + \overline{\varepsilon}  }   \nonumber   \\
    & = &     -\frac{U^2}{(2\pi^2)}
   \frac{1}{v_F \beta^{1/4}}
  \int_0^\infty dy \sqrt{\sqrt{\frac{f^2}{8\beta} + y}
- \frac{f}{\sqrt{8\beta}}} \frac{e^{-y/\Lambda}}{y} 
\label{int1}
\end{eqnarray}
where the only dependence on $\delta {\bf Q}$ comes through the
curvarture: 
$f \equiv 12\beta  \; \delta Q_x^2 - 6\alpha \;  \delta Q_x$.
A similar expression holds for the concave part of the Fermi
surface, with the same respective limit behaviors at large and 
small values of the cutoff $\Lambda$.

The result (\ref{int1})
can be rescaled as a function of the
dimensionless parameter $\frac{f^2/8\beta}{\Lambda}$, what allows
us to study the relative intensity of the corrections in
different zones of the Fermi line. We see that, close to the 
inflection points $( f \ll \Lambda )$, the vertex function
reaches maximum values 
\begin{equation}
{\rm Re} \Pi(\phi ,\phi ) \sim \frac{\Lambda^{1/4}}{\beta^{1/4}}
\label{dom}
\end{equation}
while for points with $f \gg \Lambda $
\begin{equation}
{\rm Re} \Pi(\phi ,\phi ) \sim \frac{\Lambda^{1/2}}{| f |^{1/2}}
\label{sdom}
\end{equation}
Equation (\ref{sdom}) reproduces the $2 k_F$ behavior of the
response function of an isotropic Fermi liquid.
We observe that the function $\Pi (\phi, \phi )$ has minima at
points where the value of the curvature is a maximum. Its
natural period is determined by the symmetry of the
Fermi line, which in a square lattice is in general $\pi
/2$. 

By adding $\Pi ( \phi , \phi ' )$ to the bare particle-particle
scattering, $U$,  we obtain the full irreducible vertex in the Cooper
channel. This vertex, when inserted into ladder diagrams like the ones
depicted in Fig. \ref{three}, gives the possible superconducting
instabilities. The ladder summation implicit in Fig. \ref{three} 
respects the lattice symmetries. We write the irreducible vertex as:
\begin{eqnarray}
U + \Pi (\phi_1, \phi_2 )  & = & c_0 + c_1 (\cos \phi_1  \; \cos 
   \phi_2 +  \sin \phi_1  \; \sin \phi_2 )      \nonumber      \\
  &   &   + c_2 \cos 2\phi_1 \; \cos 2\phi_2  + c_3 \sin 2\phi_1 \; 
     \sin 2\phi_2           \nonumber   \\
  &   &   + c_4 \cos 4\phi_1 \; \cos 4\phi_2  + c_5 \sin 4\phi_1
   \;   \sin 4\phi_2       +   \ldots    \label{series}
\end{eqnarray}
The effective interaction in the Cooper channel 
$\Gamma (\phi_1 , \phi_2 )$ has a similar expansion, with
coefficients $\Gamma_n$ whose
cutoff dependence is given by:
$\Gamma_n = c_n/(1 - \: c_n/(4\pi) \: \log (\Lambda /\Lambda_0)) $.
Pairing instabilities are associated to negative values of the $c_n$'s in
Eq. (\ref{series}).

Let us now assume that the expansion of $\Pi ( \phi , \phi ' )$ in
angular harmonics can be truncated at fourth order, as shown in
Eq. (\ref{series}). Then, from
Eqs. (\ref{dom}) and (\ref{sdom}), we can write: 
$\Pi (\phi, \phi ) \approx  a + b \; \sin^2 4\phi$,
where:
$a  \sim  O(\Lambda^{1/2} / f^{1/2} )$ and
$b  \sim  O(\Lambda^{1/4} / \beta^{1/4} )$, 
so that:
\begin{eqnarray}
c_0 + c_1 + c_2 + c_4 - U &  \sim  &  O(\Lambda^{1/2} / f^{1/2} ) 
        \nonumber \\
 c_3 - c_2   & \sim &  0    \nonumber \\
 c_5 - c_4   & \sim &   O(\Lambda^{1/4} / \beta^{1/4} ) 
        \label{sim1}
\end{eqnarray}
In order to close this system of equations, we compute $\Pi ( - \phi ,
\phi )$. This function is $ \sim O ( \Lambda^{1/2} / f^{1/2} )$ if
$\phi = \pi n$ and shows a crossover to a behavior $\sim O ( \Lambda )
$ for other values of $\phi$. Hence,
$\Pi (\phi, -\phi) \approx c + d \; \cos^2 \phi$,
where:
$c  \sim  O(\Lambda ) $ and
$d  \sim  O(\Lambda^{1/2} / f^{1/2})$.
It implies that:
\begin{eqnarray}
c_0 - c_1 - c_3 - c_5 - U &  \sim  &  O(\Lambda ) 
         \nonumber              \\
2 c_1   & \sim  &  O(\Lambda^{1/2} / f^{1/2})   \nonumber    \\
 c_2 + c_3   & \sim &  0   \nonumber   \\
 c_4 + c_5   & \sim &  0   \label{sim2}
\end{eqnarray}
From (\ref{sim1}) and (\ref{sim2}) we obtain that the dominant
modes in the limit $\Lambda \rightarrow 0$ 
are $c_0 - U \sim - c_4 \sim c_5 \sim O(\Lambda^{1/4} /
\beta^{1/4})$. Hence, $c_4$ is the only negative coefficient
to that leading order
and the system has an instability towards extended s-wave pairing,
with a gap which has two nodes in each quadrant of the BZ.
We stress that this argument follows the standard BCS procedure,
reformulated in RG language. The only approximation involved is the
truncation of the expansion (\ref{series}) to fourth order.

We can extend the previous analysis to anisotropic Fermi surfaces
without inflection points,
like the round shaped one shown in Fig. \ref{ffermi},
at the other side of the Van Hove filling.
Then, the vertex $\Pi ( \phi , \phi )$ is
always $\sim O ( \Lambda^{1/2} / f^{1/2})$ but it is modulated by
the local curvature of the Fermi surface.
Hence, $\Pi ( \phi , \phi  )$ has maxima for $\phi = \pi / 4$
and equivalent points, where
the curvature reaches its minimum value. The same argument gives that the
minima of $\Pi ( \phi , \phi )$ lie at $\phi = 0$ and equivalent
points. The vertex is modulated by
$\Pi (\phi, \phi ) \approx a
+ b \; \sin^2 2\phi $, where: $a \sim b \sim
O(\Lambda^{1/2} / f^{1/2})$. 
A similar analysis to that applied above leads to the
conclusion that $c_2$ is now the dominant
coefficient that becomes negative, implying a superconducting
instability with $d_{x^2 - y^2}$ symmetry, in agreement with previous
calculations for Fermi surfaces of similar shapes\cite{Schulz}.

We have assumed that the shape of the Fermi surface does not
change much upon lowering the cutoff, $\Lambda$. This is the case if
only the electrons near the Fermi surface are affected by the
interaction, that is,
$\Lambda \sim U \ll E_F$, where $E_F$ is the Fermi energy. Different
superconducting instabilities may arise in dilute systems where this
approximation is not valid\cite{kagan}.

In conclusion, we have extended the standard renormalization group
treatment of interacting electrons to systems with strongly anisotropic 
Fermi surfaces. We have shown that the existence of inflection
points change the scaling properties of the interactions. The critical
dimension which signals non-Fermi liquid behavior is 3/2.
The superconducting vertex, near the inflection points, also changes its
scaling properties. This feature allows us to propose a simple scheme to
analyze the possible superconducting instabilities of the system (the
Kohn-Luttinger mechanism). We have extended the analysis to other
anisotropic Fermi surfaces in the square lattice.
The superconducting order parameter is
determined by the modulation of the curvature around the Fermi surface.
When the curvature has two zeros in each quadrant of the BZ,
we find extended s-wave pairing. The gap has two nodes in each quadrant
of the BZ. If the curvature shows only one
minimum, we obtain d-wave
pairing, which can be $x^2 - y^2$, if the region of minimum curvature is
along the diagonal of the BZ, or of the $xy$ type, if the minimum of
curvature is along the axes. 
In all cases, the gap has a maximum in the region of maximum curvature
of the Fermi surface. This region has the lowest density of
states. It is interesting to note that a {\it screening} diagram, like
the first one in Fig. \ref{diagram}, favors a gap which spans regions
of high density of states. An unusual feature of the Kohn-Luttinger
instability is that it is induced by an {\it antiscreening} diagram, the
third in Fig. \ref{diagram}.

For the particular case of the $t - t'$ Hubbard model, we propose a
phase diagram with extended s pairing for fillings below the Van Hove
singularity, and d$_{x^2 - y^2}$ for fillings above it. At the Van Hove
filling the system shows magnetic instabilities. 
Different approximations indicate that the ferromagnetic 
instability is dominant
for a finite range of parameters\cite{ferro}.
For values of $t'$ below the ferromagnetic regime, though, 
a finite window for superconductivity
with d$_{x^2 - y^2}$ order parameter develops, as confirmed by a 
numerical diagonalization approach\cite{nd}.

The closeness of a transition towards a d-wave and an extended s-wave
superconductor found here implies that, in the absence of perfect
tetragonal symmetry, a mixture of the two is likely\cite{ds}. 
Such a possibility
is consistent with recent photoemission experiments in the overdoped
regime of BiSCCO\cite{photo}. Alternative scenarios for the interplay
between the d-wave and extended s-wave symmetry of the order parameter 
have been proposed invoking different effective
interactions\cite{dagotto,perali,abr}. 
The framework presented in this Letter,
on the other hand, links in a natural fashion the symmetry of the 
order parameter to the shape of the Fermi surface. 
Finally, our study can be relevant for the 2D anisotropic superconductor
Sr$_2$RuO$_4$\cite{Sr2RuO4exp}, which is well into the weak coupling
regime and has a superconducting state that influences a narrow energy range
near the Fermi level (see, however, Ref. \cite{Sr2RuO4th}).

\begin{figure}
\caption{Different shapes of the Fermi line for the $t-t'$
Hubbard model about the Van Hove filling. Two opposite inflection
points are marked on the figure.}
\label{ffermi}
\end{figure}

\begin{figure}
\caption{Total interaction vertex in the BCS channel.}
\label{diagram}
\end{figure}

\begin{figure}
\caption{Ladder summation in the BCS
channel used to describe the possible superconducting instabilities of
the system.}
\label{three}
\end{figure}

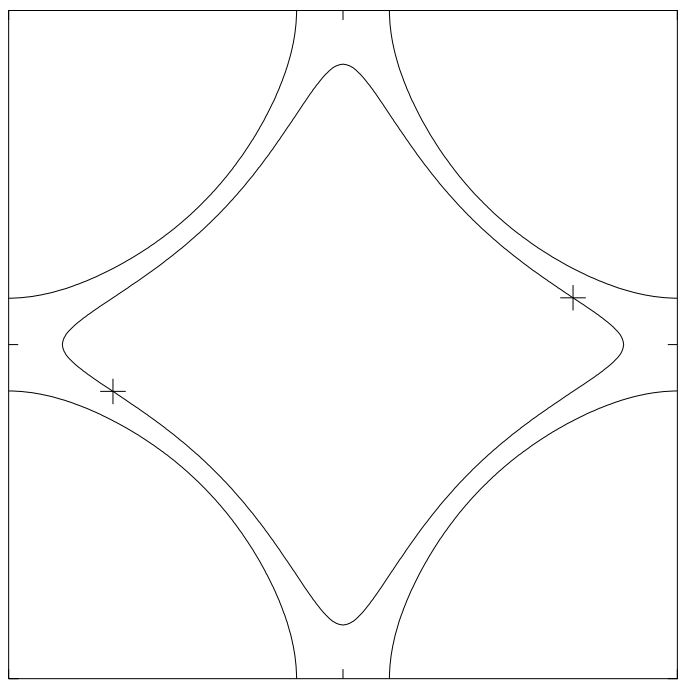
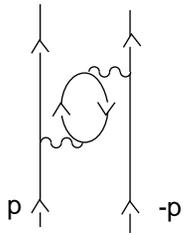
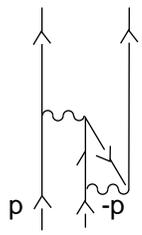
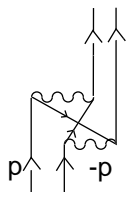
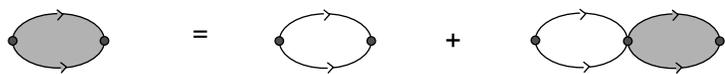
\end{document}